\documentclass[preprints,article,accept,moreauthors]{Definitions/mdpi} 

\firstpage{1} 
\pubvolume{1}
\issuenum{1}
\articlenumber{0}
\pubyear{2025}
\copyrightyear{2025}
\datereceived{ } 
\daterevised{ } 
\dateaccepted{ } 
\datepublished{ } 
\hreflink{https://doi.org/} 

\Title{Constraints on a fifth force from the stellar orbits around the central supermassive black hole of the Milky Way}

\TitleCitation{Constraints on a fifth force from the stellar orbits around the central supermassive black hole of the Milky Way}

\Author{Predrag Jovanovi\'{c}$^{1,*}$\orcidA{}, Du\v{s}ko Borka$^{2}$\orcidB{} and Vesna Borka Jovanovi\'{c}$^{2}$\orcidC{}}

\AuthorNames{Predrag Jovanovi\'{c}, Du\v{s}ko Borka and Vesna Borka Jovanovi\'{c}}

\address{%
$^{1}$ \quad Astronomical Observatory, Volgina 7, P.O. Box 74, 11060 Belgrade, Serbia \\
$^{2}$ \quad Department of Theoretical Physics and Condensed Matter Physics (020), Vin\v{c}a Institute of Nuclear Sciences - National Institute of the Republic of Serbia, University of Belgrade, P.O. Box 522, 11001 Belgrade, Serbia \\
}

\corres{Correspondence: pjovanovic@aob.rs}

\abstract{Here we investigate a possible presence of a fifth force at the Galactic Center (GC), and its potential influence on the stellar orbits around the central supermassive black hole of our Galaxy. For this purpose we simulated the stellar orbits in a Yukawa gravity model that predicts the emergence of a fifth force, and fitted them into the observed orbit of S2 star around Sgr A* at the GC. The fitting was performed using Markov chain Monte Carlo method which enabled us to constrain the parameters of Yukawa interaction describing the strength $\delta$ and the range $\lambda$ of a fifth force. We studied the following cases for a fifth force range $\lambda$, when it is: i) about a few hundred AU (i.e. deep inside the orbit of S2 star), ii) about a thousand AU (i.e. approximately the size of S2 star orbit), and iii) several thousand AU (i.e. much larger than the size of S2 star orbit). The obtained results showed that as the range $\lambda$ of a fifth force increases, its strength $\delta$ also increases and relative error $\Delta\delta/\delta$ decreases. The resulting fifth-force strengths in all three cases are respectively: $\delta\sim$ 0.005, 0.02 and 0.15. These results are consistent with the corresponding results of both our previous studies and those of other authors, regardless of the different Yukawa-like potentials used to model a fifth force. In addition, we studied whether the possible small discrepancies from the prediction of General Relativity for the Schwarzschild precession of S2 star could be also caused by a fifth force. For this purpose we used the $f_\mathrm{SP}$ parameter that was recently measured in the case of S2 star by GRAVITY Collaboration in 2020. We found that the obtained estimates in all three cases are compatible, within the error intervals, with the measured value of $f_\mathrm{SP} = 1.10\pm 0.19$.}


\begin{document}


\section{Introduction}

A fifth force is an additional effective repulsive (anti-gravity) force which could balance out the normal Newtonian attractive
gravity on galactic scales, and consequently, which could produce nearly flat rotation curves of spiral galaxies without need
for dark matter (DM) hypothesis \cite{sand84}. Moreover, since it is repulsive in its nature, a fifth force can also mimic
the dark energy (DE) effects at large scales. The concept of fifth force naturally emerges in the context of modern attempts to
unify gravity with the other fundamental interactions in nature. Specifically, it arises in the weak field limit of massive
gravity theories and some Extended Theories of Gravity (ETG) in the following form of an exponential Yukawa-like correction to
the Newtonian gravitational potential \cite{sand84,stab13,capo14}:
\begin{equation}
\Phi\left(r\right)=-\dfrac{G_{\infty}M}{r}\left(1+\delta\,e^{\displaystyle-r/\lambda}\right),\;\;\text{where}\;\;
G_{\infty}=\dfrac{G}{1+\delta}.
\label{eq:gravpot}
\end{equation}
In the above expression, $G_{\infty}$ is the gravitational constant as measured at infinity, $G$ is its local value, $\delta$ is the
coupling constant which describes the intensity (strength) of a fifth force, while $\lambda$ is a length scale representing the range of a fifth force.
The Newtonian gravitational potential can be recovered from (\ref{eq:gravpot}) in the absence of a fifth force (i.e. for $\delta=0$) in which case $G_{\infty}=G$,
or in the case when it has infinite range (i.e. for $\lambda\rightarrow\infty$). On the other
hand, as demonstrated in \cite{sand84}, the flat rotation curves of spiral galaxies could be accounted without dark matter
hypothesis in the case when $\delta\sim -1$, i.e. for a repulsive contribution of a fifth force.

Due to its potentially great importance for modern physics, there were extensive experimental attempts in the past to
constrain a fifth force on different scales. These attempts included laboratory experiments with torsion balance for $\lambda$
between $1\mu$m and $1\ \mathrm{cm}$, as well as astronomical tests by artificial satellites, Lunar Laser Ranging
and the Solar System planetary data for $\lambda$ between $1\ \mathrm{cm}$ and $10^{15}\ \mathrm{m}$, which resulted with
allowed and excluded regions for the strength and range of a fifth force on these scales (see Figs. 9 and 10 in
\cite{adel09}).

In order to explain astrophysical observations at different astronomical and cosmological scales, the modified theories of gravity have been suggested as alternative approaches to Newtonian gravity: \cite{fisc99,cope04,clif06,capo11a,capo11b,noji11,clif12,salu21,WP25}. In the context of a fifth force, the theories of massive gravity are of special importance \cite{Fierz_39,Boulware_72,Logunov_88,Chugreev_89,Gershtein_03,Gershtein_04,Gershtein_06,ruba08,babi10,deRham_11,rham14,rham17}, especially those of them with the form of Yukawa-like corrections, as described in \cite{sand84,talm88,whit01,amen04,reyn05,seal05,moff05,moff06,sere06,capo07b,iori07,iori08,capo09b,adel09,card11}, and more recently in \cite{rham17,will18,will18b,miao19,capo20,capo21,beni22b,dong22,tan24}. An important prediction of these theories, which is fundamentally different with respect to General Relativity (GR), is that gravity is propagated by a massive field (see \cite{rham14} for an extensive review). Since a fifth force arises in the weak field limit of some of these theories, its range $\lambda$ then depends on the mass $m_g$ of graviton, particle that caries
the gravitational interaction, through the expression for its Compton wavelength: $\lambda=h/(m_g\,c)$. This makes possible to
test these theories on even larger astrophysical scales and to constrain their parameters, particularly the graviton Compton
wavelength and thus, the range of a fifth force. For example, LIGO Scientific and Virgo
Collaborations constrained the graviton Compton wavelength from the first detected gravitational wave (GW) signal GW150914
from binary black holes and obtained the following estimate \cite{ligo16}: $\lambda >1.6\times 10^{13}\;\mathrm{km}$ which
corresponded to graviton mass of $m_g\le 1.2\times 10^{-22}\;\mathrm{eV}/c^2$. Recently, the LIGO - Virgo - KAGRA collaborations
updated the above bound on the mass of the graviton by studying the signals from mergers collected in The third
Gravitational-Wave Transient Catalog (GWTC-3) to $m_g\le 2.42\times 10^{-23}\;\mathrm{eV}/c^2$, at 90\% credibility \cite{ligo25}.

On the other hand, the bright S-stars which move around the Sgr A* (a compact radio source which is associated with the Supermassive black hole (SMBH) at the dynamical center of our Galaxy) represent a
powerful laboratory for studying gravity theories in such an extreme environment. S-stars have been monitored for more than 30 years \cite{ghez00,scho02,ghez08,gill09a,gill09b,genz10,meye12,gill17,hees17,chu18,abut18,abut19,do19,amor19,said19,hees20,abut20,abut22,genz22,abde25}, and a number of analyses of their orbits was performed using available observational data by several theoretical groups (see e.g. \cite{doku15,kali20,lalr21,lalr22,beni22a,beni23,bamb24}). In our recent studies  we used astrometric observations of S-stars from \cite{gill17} in order to constrain the parameters of a fifth force using the extended PPN formalism which we
developed for this purpose \cite{jova23,jova24a}. Recently, the GRAVITY Collaboration also investigated the presence of a fifth
force at the GC using Markov chain Monte Carlo (MCMC) analysis of the astrometric and spectroscopic data of S2 star \cite{abde25}.
This study resulted with a stringent upper limit for the intensity of a fifth force of $\vert\delta\vert < 0.003$ for the range
of $\lambda = 3 \times 10^{13}\ \mathrm{m}\ \sim 200\ \mathrm{AU}$ \cite{abde25}.

Slightly different forms of the Yukawa-like potential $\Phi$ were used in the aforementioned studies, depending on the specific
theory of massive gravity or ETG from which they were derived. However, the following three are the most commonly used among them for studying a fifth force:
\begin{enumerate}
\item
The phenomenological assumption in the case of massive graviton, proposed in \cite{will98}:
\begin{equation}
\Phi\left(r\right)=-\dfrac{GM}{r}\, e^{-\dfrac{r}{\lambda}},
\label{eq:pot1}
\end{equation}
which we used in our previous work \cite{jova24b} to study the stellar orbits around the GC, and which enabled us to improve our earlier constraints on the range of a fifth force and graviton mass;
\item
The potential of the form:
\begin{equation}
\Phi\left(r\right)=-\dfrac{GM}{r}\left[1+\alpha e^{-r/\lambda}\right],
\label{eq:pot2}
\end{equation}
which was recently used by the GRAVITY Collaboration for studying the presence of a fifth force at the GC \cite{abde25}, as well as
by the LIGO Scientific and Virgo Collaborations (but for strength of a fifth force of $\vert\alpha\vert=1$) for constraining the graviton Compton wavelength $\lambda$ from GW150914 \cite{ligo16};
\item
The potential:
\begin{equation}
\Phi\left(r\right)=-\dfrac{GM}{(1+\delta)r}\left({1+\delta e^{-\dfrac{r}{\lambda}}}\right),
\label{eq:yukawapot}
\end{equation}
obtained from $f(R)$ theories of gravity, which we used in our earlier papers for studying the range of Yukawa interaction $\lambda$ and its strength $\delta$ (see e.g. \cite{bork13,zakh16a,zakh18,jova21,bork21a,bork22,bork23,jova23,jova24a,jova24b,tan24,budh25}).
\end{enumerate}

Our present work represents an extension of our previous investigations in this field to the MCMC analysis of S2 star orbit,
in order to study the possible effects of a fifth force at GC, as well as to constrain its strength for different ranges $\lambda$ with respect to Sgr A*. For this purpose, we used an extension of the Parametrized Post-Newtonian (PPN) formalism \cite{clif08,alsi12,bern19,gain20a,gain20b}, which we previously developed in the case of $f(R)$ theories of gravity giving the potential (\ref{eq:yukawapot}) in the weak field approximation \cite{jova23,jova24a,jova24b}.
This Yukawa-like potential differs from those used in other recent studies of a fifth force, which enables us to compare different fifth force models and to investigate whether their choice could have a significant influence on the obtained constraints. Moreover, here we study S2 star orbit using the extended PPN formalism which differs substantially from the standard PPN formalism commonly used for modeling stellar orbits in GR and some alternative theories of gravity, as well as from the modified PPN formalism which was recently used by the GRAVITY Collaboration for studying the potential deviations of the observed stellar orbits around Sgr A* from the corresponding relativistic predictions. Furthermore, our goal here is to investigate the intensity of a fifth force and its potential effects at different distances from the Sgr A* (or more precisely, for different ranges of a fifth force relative to S2 star orbit), including the scales that have not been previously studied. Besides, here we also perform the sensitivity analysis of the obtained results in order to study their potential dependence on the assumed choice of the priors. Based on the aforementioned, we expect that this could potentially lead to new physical insights about a fifth force, as well as to improve constraints on its strength and range.

This paper is organized as follows: in Section 2 we describe our method based on extended PPN formalism and MCMC analysis, in Section 3 we present and discuss our obtained results, and finally we devote Section 4 to our concluding remarks.

\section{Constraining a fifth force by the stellar orbits around Sgr A*}

\subsection{Extended PPN formalism}

In our recent papers \cite{jova23,jova24a,jova24b}, we used the so called extended PPN formalism to calculate a simulated true orbit of S2 star in Yukawa gravity. The true orbit is then projected onto the observer's sky plane in order to obtain the corresponding apparent orbit (for more details see \cite{jova24a}). In this extended PPN formalism the equation of motion in Yukawa gravity is given by \cite{dong22}:
\begin{equation}
\vec{\ddot{r}}=-GM \dfrac{\vec{r}}{r^3}+
\dfrac{GM}{c^2r^3} \left[\left(4\dfrac{G M}{r}-\vec{\dot{r}}\cdot\vec{\dot{r}}\right) \vec{r} + 4\left(\vec{r}\cdot\vec{\dot{r}}\right)\vec{\dot{r}}\right]+
\dfrac{\delta \cdot GM}{1 + \delta} \left[ 1 - \left(1 + \dfrac{r}{\lambda} \right) e^{-\dfrac{r}{\lambda}} \right] \dfrac{\vec{r}}{r^3}.
\label{eq:ppneom}
\end{equation}
The last term in r.h.s is an addition to the standard PPN formalism for equation of motion in GR (see \cite{will18} for more details), and it vanishes when $\delta=0$ or $\lambda\rightarrow\infty$, recovering  the standard PPN equation of motion. This modification arises because the standard PPN formalism is not viable for Yukawa gravity and requires an additional Yukawa-like correction.

In our previous studies \cite{jova23,jova24a,jova24b}, we also derived the following condition which should be satisfied by the
strength $\delta$ and range $\lambda$ of a fifth force, in order that the orbital precession in the extended PPN formalism (\ref{eq:ppneom})
is close to the observed precession of S2 star obtained by the GRAVITY Collaboration \cite{abut20}:
\begin{equation}
\lambda(P,e,\delta,f_\mathrm{SP}) \approx \dfrac{cP}{2\pi} \sqrt{\dfrac{\delta (\sqrt{1-e^2})^3}{6(f_\mathrm{SP} - 1)(1 + \delta)}},
\label{eq:lambda}
\end{equation}
where $P$ is orbital period, $e$ is eccentricity and $a$ is semi-major axis of S2 star orbit. This approximation is valid only when $\lambda(P,e,\delta,f_\mathrm{SP}) \gg a$. Absolute error of the $\lambda$ can be estimated using its total derivative:
\begin{equation}
\Delta\lambda\approx\left|\dfrac{\partial\lambda}{\partial P}\right|\Delta P+\left|\dfrac{\partial\lambda}{\partial e}\right|\Delta e+
\left|\dfrac{\partial\lambda}{\partial\delta}\right|\Delta\delta+\left|\dfrac{\partial\lambda}{\partial f_\mathrm{SP}}\right|\Delta f_\mathrm{SP},
\label{eq:totalderiv}
\end{equation}
which results with (see e.g. \cite{jova24a}):
\begin{equation}
\Delta\lambda\approx\lambda\cdot\left(\dfrac{\Delta P}{P}+\dfrac{3 e\Delta e}{2 (1-e^2)}+
\dfrac{\Delta\delta}{2\left|\delta\left(\delta+1\right)\right|}+\dfrac{\Delta f_{SP}}{2\left| f_{SP}-1\right|}\right).
\label{eq:lambdaerror}
\end{equation}

The parameter $f_\mathrm{SP}$ in above expressions (\ref{eq:lambda})-(\ref{eq:lambdaerror}) was introduced by the GRAVITY Collaboration \cite{abut20} in order to parametrize the effect of the Schwarzschild metric and to study to which extent some gravitational model is relativistic. Theoretically, it is equal to $f_{SP} = (2 + 2\gamma - \beta) / 3$, where $\beta$ and $\gamma$ are the standard post-Newtonian parameters. In the case of GR, $\beta$ and $\gamma$ are both equal to 1, and thus $f_{SP}=1$. Therefore, here we will use the expressions (\ref{eq:lambda}) and (\ref{eq:lambdaerror}) to study whether the best-fit orbits of S2 star, obtained in the presence of a fifth force, are compatible with the value of $f_{SP} = 1.10\pm 0.19$ which was measured by GRAVITY Collaboration \cite{abut20}.

\subsection{MCMC analysis of S2 star orbit and a fifth force parameters}

In order to obtain the estimates of strength $\delta$ and range $\lambda$ of a fifth force at GC, we calculated the simulated orbits of S2 star in the extended PPN formalism (\ref{eq:ppneom}) and fitted them into the astrometric observations of S2 star from \cite{gill17}. The best-fit values and uncertainties of S2 star orbital elements, as well as of strength and range of a fifth force are obtained using the Markov chain Monte Carlo method.

MCMC analysis provides very efficient sampling approximations to the posterior probability density function in parameter spaces \cite{hogg18}. At the beginning of a MCMC analysis it is necessary to define the likelihood function $\cal{L}$ and priors which encode any previous knowledge about the parameters. As a result of a such analysis, one obtains an estimate of the posterior probability distributions of the parameters which show all the covariances between them, as well as to what extent they are consistent with the observed dataset. The likelihood function $\cal{L}$ is the probability of a dataset given the model parameters, or in other words, it measures how well a statistical model explains the observed data. In most cases, it is proportional to $\exp(-\chi^2/2)$, where $\chi^2$ statistic measures the difference between the data and a model's prediction. Thus, maximizing likelihood $\cal{L}$ is equivalent to minimizing $\chi^2$.

For the orbital fitting, we used \href{https://emcee.readthedocs.io/en/stable/}{\nolinkurl{emcee}} - the Python implementation of Goodman \& Weare's Affine Invariant MCMC Ensemble sampler \cite{fore13}. This MCMC implementation uses the logarithm of the likelihood function (log-likelihood), which is thus given by $\ln{\cal{L}}=-\chi^2/2$ for a single set of the observational data. However, in the case of S2 star separate astrometric and radial velocity observational data sets are available, and they can be included into the total log-likelihood by taking the sum of their individual log-likelihoods: $\ln{\cal{L}}=\ln{\cal{L}}_\mathrm{pos}+\ln{\cal{L}}_\mathrm{vel}$, where $\ln{\cal{L}}_\mathrm{pos}=-\chi^2_\mathrm{pos}/2$ and $\ln{\cal{L}}_\mathrm{vel}=-\chi^2_\mathrm{vel}/2$ (see also \cite{abde25}). Given that radial velocity data from \cite{gill17} have sparse cadence and covers much shorter observational period than the corresponding astrometric data, we used only the latter for fitting the S2 star orbit. The fitted radial velocities are then calculated from the positions and velocities along the resulting best-fit orbit, according to (see \cite{bork13} for more details):
\begin{equation}
v_\mathrm{rad} = \dfrac{\sin i}{\sqrt{x^2 + y^2}} \left[ \sin(\theta + \omega) \cdot(x \dot{x} + y \dot{y}) + \cos (\theta + \omega) \cdot (x \dot{y} - y \dot{x})\right], \quad \theta = \arctan \dfrac{y}{x}.
\label{vrad}
\end{equation}
Thus, in our analysis the reduced $\chi^2$ between the observed $(x_i^o, y_i^o)$ and calculated $(x_i^c, y_i^c)$ positions of S2 star along its orbit is given by:
\begin{equation}
\chi^2 = \chi^2_\mathrm{pos} = \dfrac{\chi_x^2+\chi_y^2}{2}=\dfrac{1}{2\left(N-\nu\right)}{\sum\limits_{i = 1}^N {\left[ {{{\left( {\dfrac{x_i^o - x_i^c}{\sigma_{xi}}}
\right)}^2} + {{\left( \dfrac{y_i^o - y_i^c}{\sigma_{yi}} \right)}^2}} \right]} },
\label{eq:chi2}
\end{equation}
where $N$ is the number of observations, $\nu$ is the number of unknown parameters and $(\sigma_{xi}, \sigma_{yi})$ are the observed astrometric uncertainties.

We should note here that our goal was not to study the distance $R_0$ to Sgr A* or its offset with respect to the reference frame, for which there are very robust estimates obtained from more recent observations \cite{abde24}. Therefore, we chose to fit only the minimum number of parameters which are necessary for constraining a fifth force at GC. Hence, our set of parameters consisted of the SMBH mass $M$, 6 orbital elements of S2 star: $a$[arcsec] - semi-major axis, $e$ - eccentricity, $i$[$^\circ$] - inclination, $\Omega$[$^\circ$] - longitude of the ascending node, $\omega$[$^\circ$] - pericenter argument and $T$[yr] - epoch of the pericenter passage, as well as of 2 parameters describing a fifth force: $\delta$ - its strength, and $\lambda$[AU] - its range.

\begin{table}[ht!]
\centering
\caption[]{Priors for the MCMC analysis which are common for all studied cases of S2 star orbit.}
\label{tab1}
\begin{tabular}{l|lll|ll}
\hline
\noalign{\smallskip}
 & \multicolumn{3}{c|}{Uniform priors} & \multicolumn{2}{c}{Gaussian priors} \\
\noalign{\smallskip}
\cline{2-6}
Parameter& Initial value & Lower bound & Upper bound & $\mu$ & $\sigma$ \\
\noalign{\smallskip}
\hline
\noalign{\smallskip}
$M\ (\times 10^6\ M_\odot)$& 4 & 0.1 & 8 & 4.28 & 0.1 \\
$a\ ('')$                  & 0.1255 & 0.10 & 0.15 & 0.1255 & 0.0009 \\
$e$                        & 0.8839 & 0.78 & 0.99 & 0.8839 & 0.0019 \\
$i\ (^\circ)$              & 134.18 & 90 & 150 & 134.18 & 0.40 \\
$\Omega\ (^\circ)$         & 226.94 & 190 & 250 & 226.94 & 0.60 \\
$\omega\ (^\circ)$         & 65.51 & 20 & 90 & 65.51 & 0.57 \\
$T$ (yr)                   & 2002 & 2001 & 2004 & 2002.33 & 0.01 \\
\noalign{\smallskip}
\hline
\noalign{\smallskip} \noalign{\smallskip}
\end{tabular}
\end{table}

\begin{table}[ht!]
\centering
\caption[]{Specific uniform priors for strength $\delta$ and range $\lambda$ of a fifth force in three studied cases.}
\label{tab2}
\begin{tabular}{lllll}
\hline
\noalign{\smallskip}
Case & Parameter& Initial value & Lower bound  & Upper bound \\
\noalign{\smallskip}
\hline
\noalign{\smallskip}
1&$\delta$      & $1 \times 10^{-3}$ & -0.2 & 0.2 \\
&$\lambda$ (AU) & 250 & 200 & 500 \\
\noalign{\smallskip}
\hline
\noalign{\smallskip}
2&$\delta$      & $1 \times 10^{-3}$ & -0.3 & 0.3 \\
&$\lambda$ (AU) & $2 \times 10^3$ & 500 & $2.5 \times 10^3$ \\
\noalign{\smallskip}
\hline
\noalign{\smallskip}
3&$\delta$      & 0.2 & -0.95 & 0.95 \\
&$\lambda$ (AU) & $5 \times 10^3$ & $1 \times 10^3$ & $1 \times 10^4$ \\
\noalign{\smallskip}
\hline
\noalign{\smallskip} \noalign{\smallskip}
\end{tabular}
\end{table}

We used uniform priors for all fitted parameters which are listed in Tables \ref{tab1} and \ref{tab2}. The priors for the SMBH mass and S2 star orbital elements from Table \ref{tab1} are common for all studied cases, while Table \ref{tab2} contains the priors for strength $\delta$ and range $\lambda$ of a fifth force which are specific for each of three studied cases. The initial values for S2 star orbital elements in Table \ref{tab1} are taken from Table 3 in Appendix D of \cite{gill17}. The numerical integration of the equation of motion in the extended PPN formalism (\ref{eq:ppneom}) is performed using the SciPy function \href{https://docs.scipy.org/doc/scipy/reference/generated/scipy.integrate.solve_ivp.html}{\nolinkurl{scipy.integrate.solve_ivp}} with a Runge-Kutta 5(4) algorithm. Orbit integration is performed starting from the initial conditions described by the state vector consisting of two position coordinates and two velocity components at the epoch $t_0$ of the first observation. These initial conditions were obtained from the Keplerian orbit, by solving Kepler's equation at epoch $t_0$ using the Newton-Raphson method implemented in the SciPy function \href{https://docs.scipy.org/doc/scipy/reference/generated/scipy.optimize.newton.html}{\nolinkurl{scipy.optimize.newton}}.

\begin{table}[ht!]
\centering
\caption[]{Results of the MCMC analysis with the uniform priors., given for three chosen cases. Best-fit values of the parameters and their uncertainties are obtained from the 16th, 50th and 84th percentiles of the posterior probability distributions.}
\label{tab3}
\begin{tabular}{p{2.5cm}p{2.5cm}p{2.5cm}p{2.3cm}}
\hline
\noalign{\smallskip}
Parameter & Case 1 & Case 2 & Case 3 \\
\noalign{\smallskip}
\hline
\noalign{\smallskip}
$M\ (\times 10^6\ M_\odot)$ & $3.89^{+1.10}_{-0.70}$       & $3.97^{+1.16}_{-0.76}$        & $3.98^{+1.13}_{-0.73}$          \\
\noalign{\smallskip}
$a\ ('')$                   & $0.1207^{+0.0102}_{-0.0074}$ & $0.1215^{+0.0106}_{-0.0078}$  & $0.1217^{+0.0103}_{-0.0074}$    \\
\noalign{\smallskip}
$e$                         & $0.8704^{+0.0266}_{-0.0310}$ & $0.8739^{+0.0255}_{-0.0286}$  & $0.8743^{+0.0246}_{-0.0283}$    \\
\noalign{\smallskip}
$i\ (^\circ)$               & $134.93^{+4.13}_{-3.56}$     & $134.85^{+4.35}_{-3.88}$      & $134.95^{+4.26}_{-4.09}$        \\
\noalign{\smallskip}
$\Omega\ (^\circ)$          & $221.85^{+10.95}_{-11.20}$   & $223.20^{+9.81}_{-10.42}$     & $223.31^{+9.29}_{-9.89}$        \\
\noalign{\smallskip}
$\omega\ (^\circ)$          & $60.24^{+11.97}_{-14.08}$    & $61.61^{+10.20}_{-11.75}$     & $61.96^{+8.76}_{-10.77}$        \\
\noalign{\smallskip}
$T$ (yr)                    & $2002.16^{+0.31}_{-0.36}$    & $2002.20^{+0.27}_{-0.32}$     & $2002.21^{+0.26}_{-0.29}$       \\
\noalign{\smallskip}
$\delta$                    & $0.0052^{+0.0625}_{-0.0547}$ & $0.0249^{+0.1637}_{-0.1516}$  & $0.1470^{+0.5297}_{-0.5371}$    \\
\noalign{\smallskip}
$\lambda$ (AU)              & $357.42^{+99.24}_{-103.27}$  & $1765.55^{+522.40}_{-711.42}$ & $6021.47^{+2700.16}_{-3053.74}$ \\
\noalign{\smallskip}
\hline
\noalign{\smallskip} \noalign{\smallskip}
\end{tabular}
\end{table}

\begin{table}[ht!]
\centering
\caption[]{The same as in Table \ref{tab3}, but for the Gaussian priors.}
\label{tab4}
\begin{tabular}{p{2.5cm}p{2.5cm}p{2.5cm}p{2.3cm}}
\hline
\noalign{\smallskip}
Parameter & Case 1 & Case 2 & Case 3 \\
\noalign{\smallskip}
\hline
\noalign{\smallskip}
$M\ (\times 10^6\ M_\odot)$ & $4.35^{+0.06}_{-0.06}$       & $4.34^{+0.07}_{-0.07}$        & $4.35^{+0.07}_{-0.07}$          \\
\noalign{\smallskip}
$a\ ('')$                   & $0.1251^{+0.0005}_{-0.0005}$ & $0.1251^{+0.0005}_{-0.0005}$  & $0.1251^{+0.0005}_{-0.0005}$    \\
\noalign{\smallskip}
$e$                         & $0.8840^{+0.0017}_{-0.0017}$ & $0.8839^{+0.0017}_{-0.0017}$  & $0.8840^{+0.0017}_{-0.0017}$    \\
\noalign{\smallskip}
$i\ (^\circ)$               & $134.24^{+0.32}_{-0.32}$     & $134.24^{+0.32}_{-0.32}$      & $134.26^{+0.32}_{-0.32}$        \\
\noalign{\smallskip}
$\Omega\ (^\circ)$          & $226.78^{+0.53}_{-0.51}$     & $226.78^{+0.50}_{-0.51}$      & $226.87^{+0.51}_{-0.52}$        \\
\noalign{\smallskip}
$\omega\ (^\circ)$          & $65.65^{+0.53}_{-0.54}$      & $65.64^{+0.53}_{-0.53}$       & $65.47^{+0.54}_{-0.51}$         \\
\noalign{\smallskip}
$T$ (yr)                    & $2002.33^{+0.01}_{-0.01}$    & $2002.33^{+0.01}_{-0.01}$     & $2002.33^{+0.01}_{-0.01}$       \\
\noalign{\smallskip}
$\delta$                    & $-0.0069^{+0.0071}_{-0.0072}$ & $-0.0281^{+0.0288}_{-0.0388}$ & $-0.1136^{+0.2966}_{-0.2270}$  \\
\noalign{\smallskip}
$\lambda$ (AU)              & $362.16^{+95.86}_{-107.23}$  & $1902.31^{+432.16}_{-703.76}$ & $7399.83^{+1846.27}_{-2766.70}$ \\
\noalign{\smallskip}
\hline
\noalign{\smallskip} \noalign{\smallskip}
\end{tabular}
\end{table}

\section{Results and discussion}

In order to investigate the effects of a fifth force on S2 star orbit around Sgr A*, we analyzed the following three cases when a fifth force range $\lambda$ is:
\begin{enumerate}
\item\hspace{-1.5em}
about a few hundred AU (i.e. deep inside the orbit of S2 star)
\item\hspace{-1.5em}
about a thousand AU (i.e. approximately the size of S2 star orbit), and
\item\hspace{-1.5em}
several thousand AU (i.e. much larger than the size of S2 star orbit).
\end{enumerate}
For each of these cases we found the best-fit orbit of S2 star by MCMC analysis, using the corresponding priors for $\delta$ and $\lambda$ from Table \ref{tab2}.

\begin{figure}[ht!]
\centering
\includegraphics[width=0.66\textwidth]{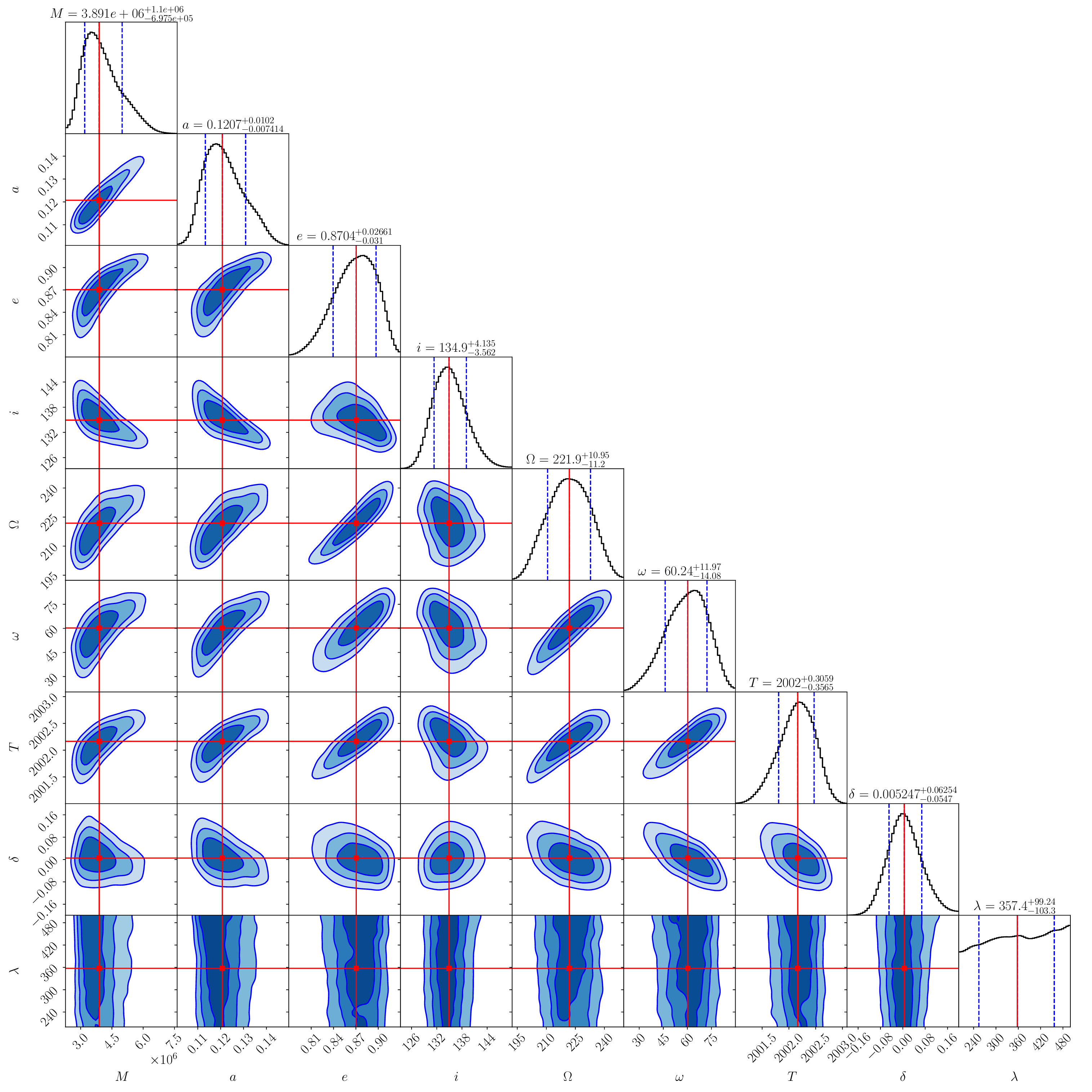}
\includegraphics[width=0.66\textwidth]{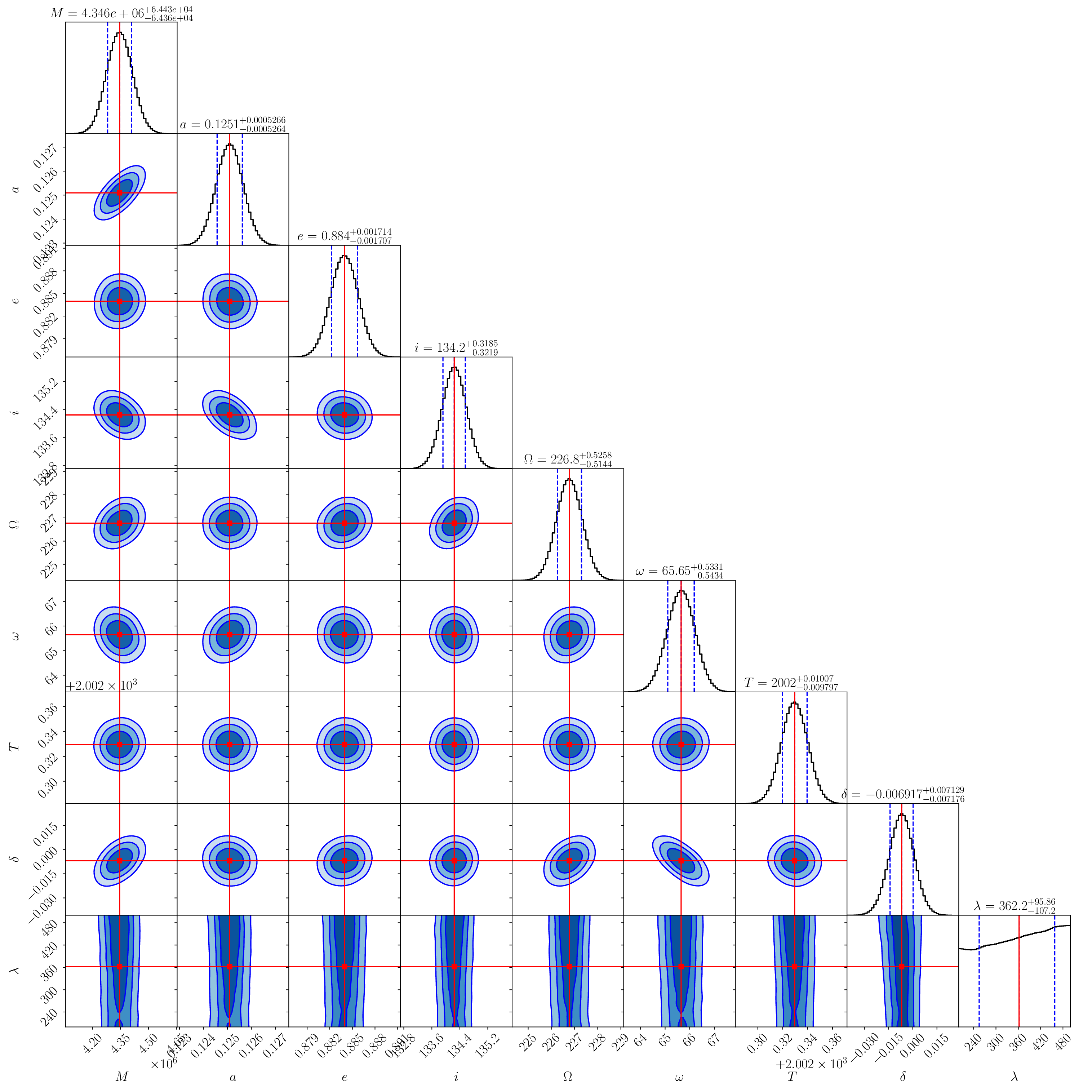}
\caption{Posterior distribution of S2 star orbit fit in the case 1 from Table \ref{tab3}, as determined from MCMC analysis.
Blue shaded 2D contours represent the confidence levels of 39.3\%, 67.5\% and 86.4\% which correspond to the $1\sigma$, $1.5 \sigma$ and $2 \sigma$ confidence regions for a 2D Gaussian distribution. The best-fit values of the parameters (red cross over 2D contours), as well as their uncertainties, were obtained from the 16th, 50th and 84th percentiles (vertical lines in 1D histograms) of the corresponding posterior probability distributions. The results for the uniform priors are shown in the top panel, while the corresponding results for the Gaussian priors are presented in the bottom panel.}
\label{fig1}
\end{figure}

\begin{figure}[ht!]
\centering
\includegraphics[width=0.75\textwidth]{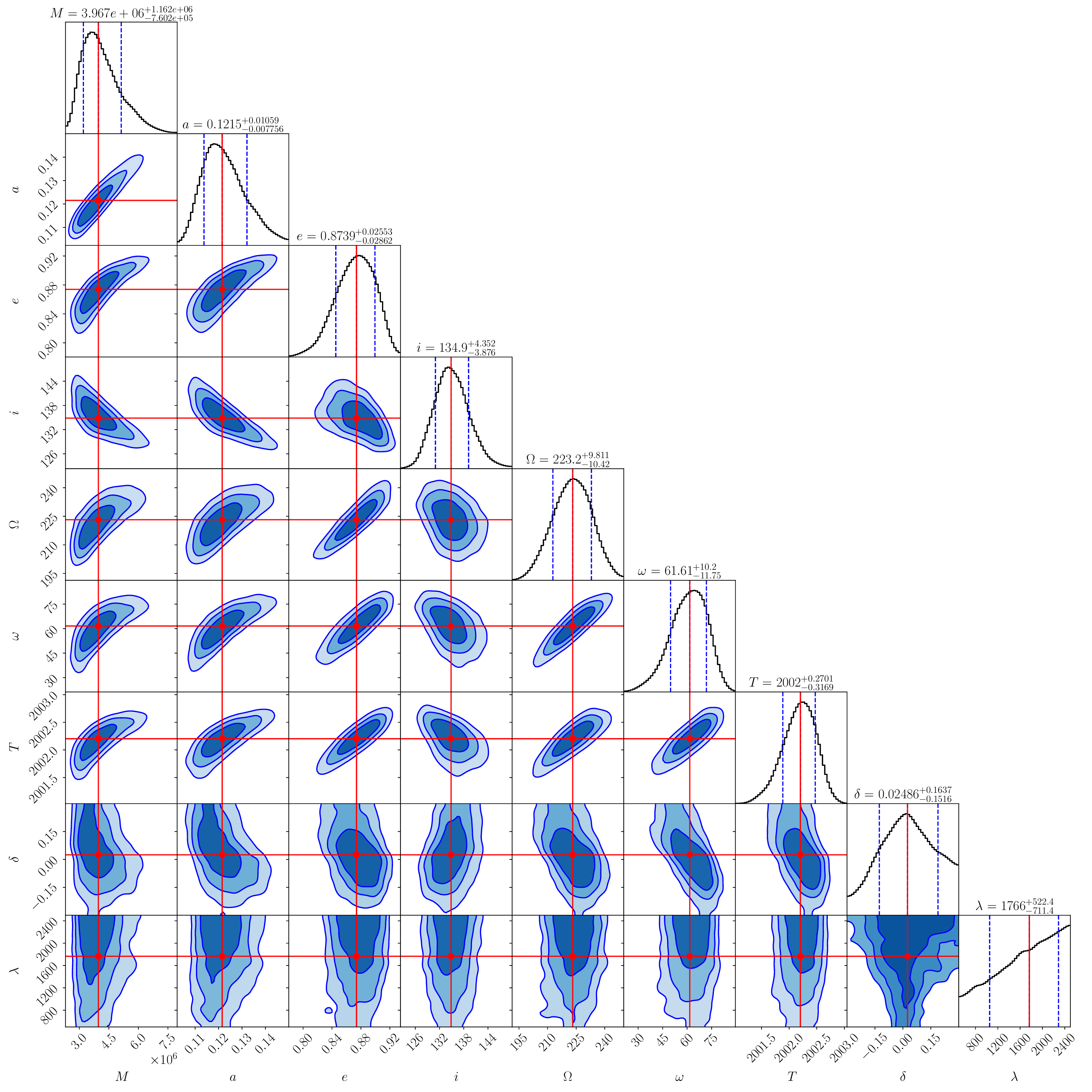}
\includegraphics[width=0.75\textwidth]{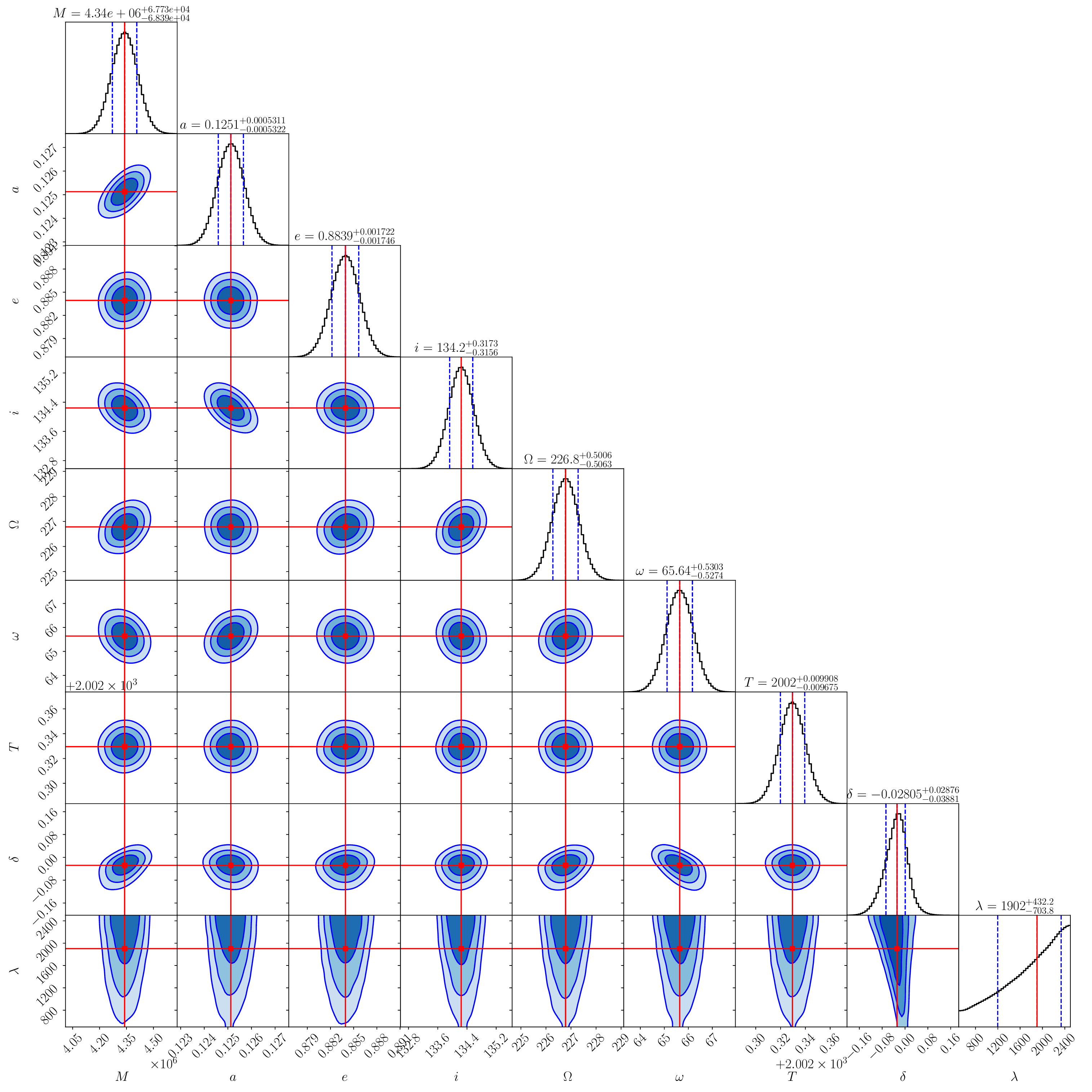}
\caption{The same as in Fig. \ref{fig1}, but for case 2 from Table \ref{tab2}.}
\label{fig2}
\end{figure}

\begin{figure}[ht!]
\centering
\includegraphics[width=0.75\textwidth]{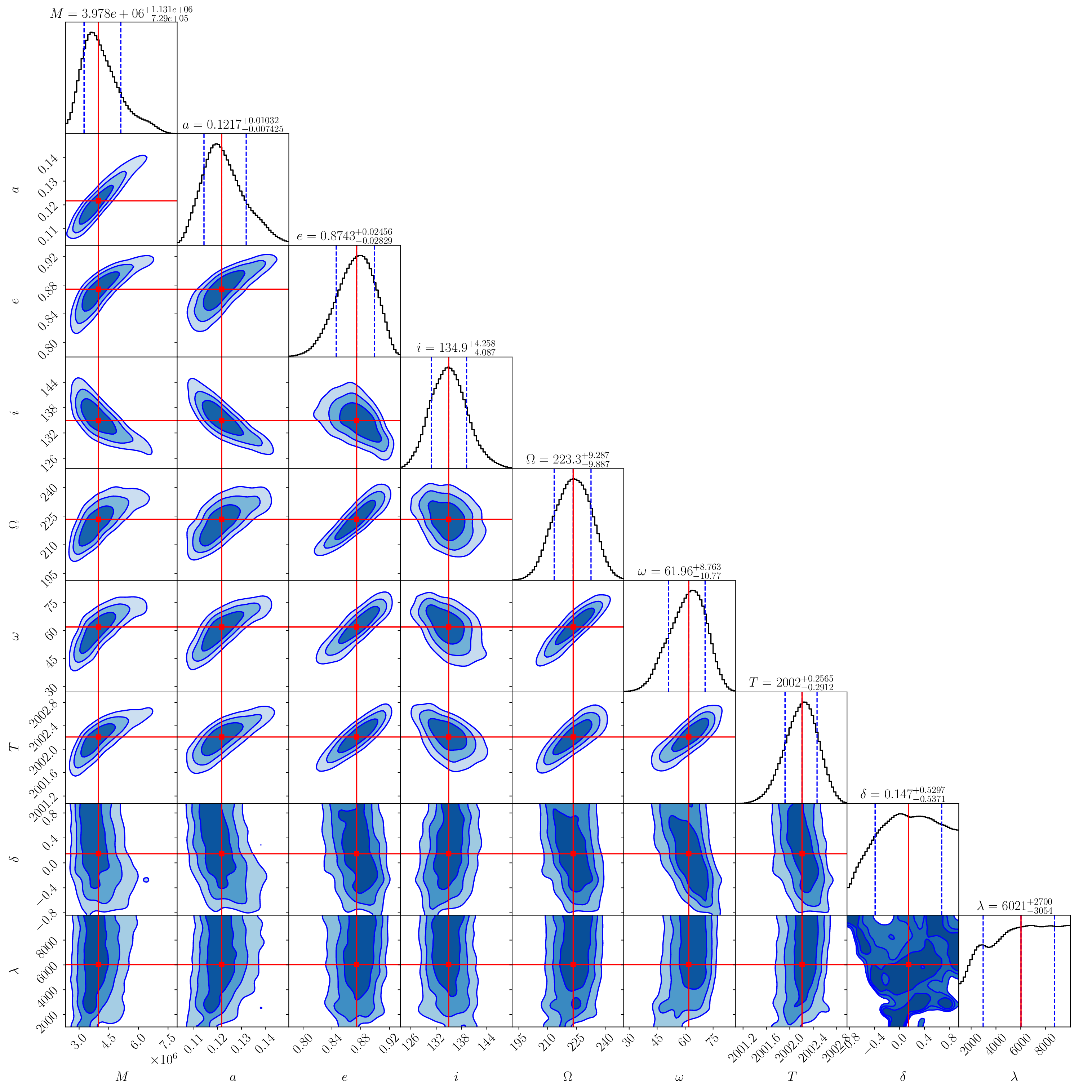}
\includegraphics[width=0.75\textwidth]{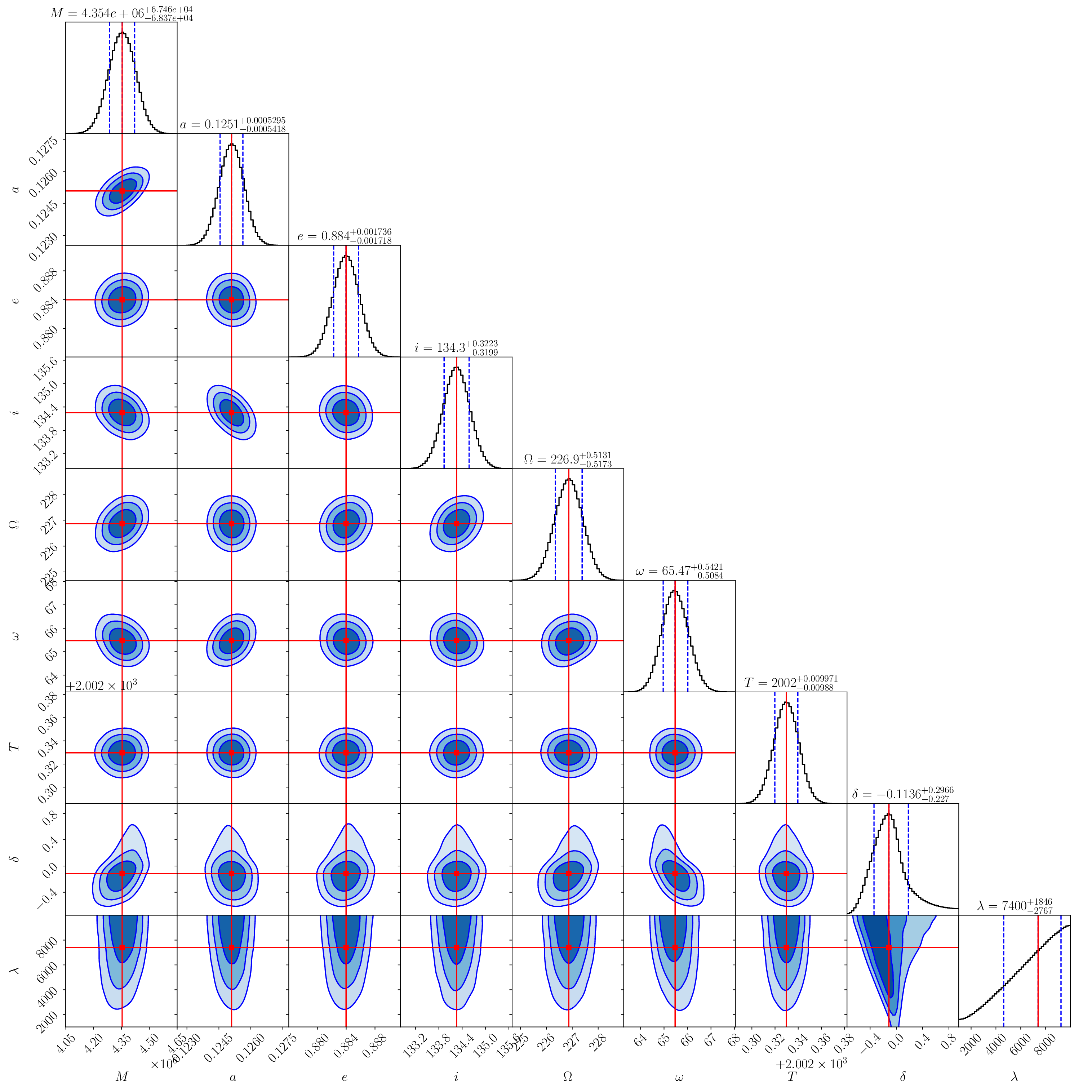}
\caption{The same as in Fig. \ref{fig1}, but for case 3 from Table \ref{tab2}.}
\label{fig3}
\end{figure}

\begin{figure}[ht!]
\centering
\includegraphics[width=\textwidth]{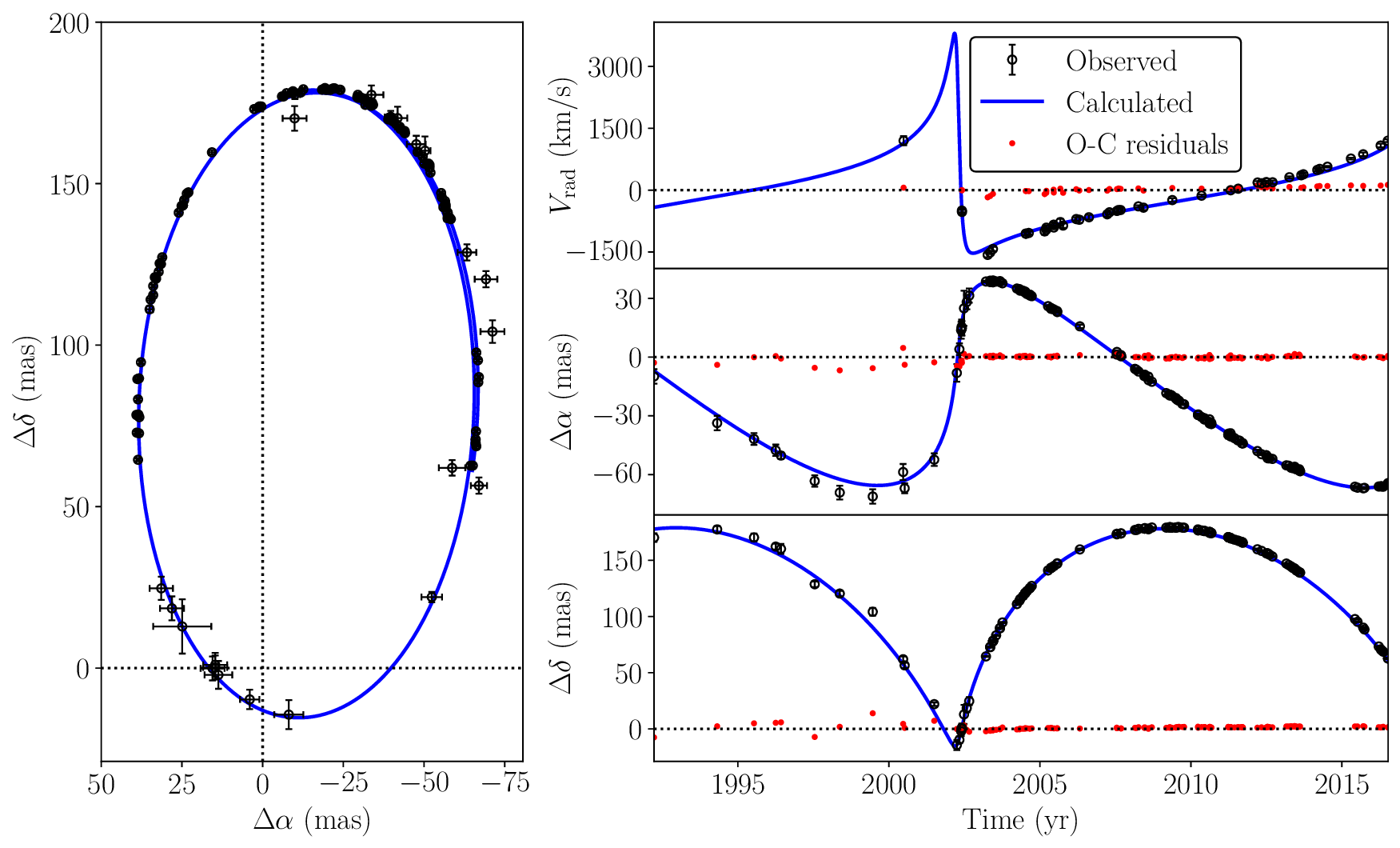}
\caption{\textit{Left:} Comparison between the best-fit orbit of the S2 star (blue solid line), obtained for the uniform priors from Table \ref{tab1} in the case 1 from Table \ref{tab2}, and the corresponding astrometric observations from \cite{gill17} (black circles with error bars).
\textit{Right:} The same for the radial velocity of the S2 star (top), as well as for its $\alpha$ (middle) and $\delta$
(bottom) offset relative to the position of Sgr A* at the coordinate origin. Red dots in the right panels denote the
corresponding O-C residuals.}
\label{fig4}
\end{figure}

The obtained results of our MCMC analyses for all three studied cases are presented in Figs. \ref{fig1}-\ref{fig4} and Tables \ref{tab3} and \ref{tab4}. Figs. \ref{fig1}-\ref{fig3} show the obtained posterior probability distributions of the fitted parameters, while Fig. \ref{fig4} presents an example of the best-fit orbit of S2 star, obtained in the first studied case. Orbital fits in the other two cases are omitted due to their visual similarity with the presented one. The corresponding numerical results for the best-fit values and uncertainties of the parameters, obtained from the 16th, 50th and 84th percentiles of their posterior probability distributions, are given in Tables \ref{tab3} and \ref{tab4}.

In our three studied cases, for the parameters from Table \ref{tab1} we used the same prior values, and for $\delta$ and $\lambda$ we chose the different prior ranges (Table \ref{tab2}) in order to investigate the effects of fifth force at different scales. We first performed MCMC analysis using the uniform priors for the parameters from Table \ref{tab1}, since they have wider distribution than the Gaussian priors and allow greater variability of the parameters, which is convenient for probing wider range of possibilities. As noted in \cite{hass26}, their prior distributions chosen for all parameters in the Milky Way were uniform, as the authors wanted to probe a wide range of possibilities, thus avoiding too narrow distributions. Also, as explained in the supplement material of \cite{do19}, the authors performed orbital fitting using the uniform priors for all parameters. In Ref. \cite{nava26} the authors assume uniform prior distributions for the orbital parameters, and for gravitational parameters, too. It can be found in the literature that even the combination of uniform (flat) and Gaussian priors can be used. For instance, in the paper of \cite{bamb26}, the authors clearly showed that the uniform priors for all parameters, except the reference frame offsets, were used. Also, in Ref. \cite{abde25} the uniform priors were used for the physical parameters, and Gaussian priors for the offset parameters. Given all of that and similarly to e.g. \cite{qi26}, we also performed an additional MCMC analysis using the Gaussian priors, also listed in Table \ref{tab1} (4th and 5th column), in order to check to what extent the obtained results are sensitive to the choice of the priors.

\subsection{Uniform priors}

As it can be seen from the top panels of Figs. \ref{fig1}-\ref{fig3} and Table \ref{tab3} which contain the results for the uniform priors, the obtained fifth-force strengths in all three cases are: $\delta\sim$ 0.005, 0.02 and 0.15, respectively. This could indicate that the strength $\delta$ of a fifth force increases with the increase in its range $\lambda$. The value $\delta\approx 0.005$, obtained in the first case, is in good agreement with the recent constraints on the strength and range of a fifth force obtained by the GRAVITY Collaboration \cite{abde25} for a similar length scale, but for a different Yukawa-like potential. This comparison shows that the strength of a fifth force is almost model independent and that different Yukawa-like potentials could be used for modeling a fifth force. Moreover, the results obtained in the third case are compatible with our earliest estimates for Yukawa gravity parameters given in \cite{bork13}.

Concerning the above results, one should have in mind that the obtained uncertainties of $\delta$ are large, probably due to insufficient accuracy of the used astrometric observations of S2 star, and that overcoming this issue would require future, more precise observations.

\subsection{Gaussian priors and sensitivity analysis}

The corresponding results for the Gaussian priors are presented in the bottom panels of Figs. \ref{fig1}-\ref{fig3} and Table \ref{tab4}, from which it can be seen that the obtained posterior distributions are much narrower than those for the uniform priors, which results with significantly smaller uncertainties of all parameters. Moreover, the Gaussian priors enforce Gaussianity in the resulting posterior distributions, in contrast to the uniform priors, for which the resulting posterior distributions are sometimes complex and non-Gaussian, as it is the case for e.g. SMBH mass $M$. Such difference in the shapes of the posterior distributions is the most likely cause of the lower values of $M$ obtained for the uniform priors compared to the corresponding values for the Gaussian priors, and may indicate the potential dependence of this parameter on the choice of the priors. For most of the remaining parameters from Tables \ref{tab3} and \ref{tab4}, except for a fifth-force strength $\delta$, the similar results are obtained for both uniform and Gaussian priors, indicating their independence on the prior choice. Regarding a fifth-force strength, the obtained values in all three cases are of the same order of magnitude as for the uniform priors, but with negative sign: $\delta\sim$ -0.007, -0.03 and -0.11, respectively. By considering its absolute values, a similar conclusion about model independence of a fifth force and increase of its strength $\delta$ with the increase in its range $\lambda$ could be drawn as for uniform priors. Since for both uniform and Gaussian priors the obtained values of $\delta$ are always smaller than its uncertainties, it is not possible to draw a solid conclusion about the dependence of this parameter on the choice of the priors.

Due to above mentioned non-Gaussianity of the posterior distributions for the uniform priors, a direct comparison of the results from Tables \ref{tab3} and \ref{tab4} may not be sufficiently robust method for testing the dependence of the obtained results on the prior assumptions. Therefore, we performed an additional sensitivity analysis by comparing the mean values of the parameters for the uniform and Gaussian priors, obtained in all three studied cases by averaging the parameters over all posterior samples remaining after discarding burn-in steps in order to ensure convergence. Such averages should be significantly less affected by the non-Gaussianity than the medians, given in Tables \ref{tab3} and \ref{tab4}, that are obtained from the 50th percentiles of the posterior distributions, and thus they should represent better indicators for the potential dependence of the parameters on the choice of the priors. The mean values of the parameters obtained for both uniform and Gaussian priors are presented in Table \ref{tab5}, from which it can be seen that the differences in masses $M$ for these two type of priors are significantly reduced, but still present, as well is the negative sign of a fifth force strength $\delta$. As a summary, this analysis shows that there is a certain dependence of the SMBH mass $M$ and a fifth force parameters $\delta$ and $\lambda$ on the choice of the priors that may somewhat affect the statistical significance of the corresponding results, which should therefore not be taken as particularly robust.

\begin{table}[ht!]
\centering
\caption[]{Comparisons between the mean values of the parameters resulting from the uniform and Gaussian priors in all three studied cases. The presented means are obtained by averaging over all posterior samples remaining after discarding burn-in steps to ensure convergence.}
\label{tab5}
\begin{tabular}{l|l|l|l|l|l|l}
\hline
\noalign{\smallskip}
 & \multicolumn{2}{c|}{Case 1} & \multicolumn{2}{c|}{Case 2} & \multicolumn{2}{c}{Case 3} \\
\noalign{\smallskip}
\cline{2-7}
\noalign{\smallskip}
Parameter & Uniform & Gaussian & Uniform & Gaussian & Uniform & Gaussian \\
\noalign{\smallskip}
\hline
\noalign{\smallskip}
$M\ (\times 10^6\ M_\odot)$ & 4.05       & 4.35        & 4.15      & 4.34       & 4.16     & 4.35       \\
$a\ ('')$                   & 0.1218     & 0.1251      & 0.1227    & 0.1251     & 0.1229   & 0.1251     \\
$e$                         & 0.8682     & 0.8839      & 0.8721    & 0.8839     & 0.8725   & 0.8840     \\
$i\ (^\circ)$               & 135.22     & 134.24      & 135.14    & 134.24     & 135.11   & 134.26     \\
$\Omega\ (^\circ)$          & 221.77     & 226.78      & 222.98    & 226.78     & 223.11   & 226.87     \\
$\omega\ (^\circ)$          & 59.32      & 65.65       & 60.77     & 65.64      & 61.22    & 65.48      \\
$T$ (yr)                    & 2002.14    & 2002.33     & 2002.18   & 2002.33    & 2002.19  & 2002.33    \\
$\delta$                    & 0.0090    & -0.0069      & 0.0262    & -0.0313    & 0.1386   & -0.0710    \\
$\lambda$ (AU)              & 355.88     & 357.97      & 1693.21   & 1795.59    & 5900.33  & 7010.34    \\
\noalign{\smallskip}
\hline
\noalign{\smallskip} \noalign{\smallskip}
\end{tabular}
\end{table}

\subsection{Posterior correlations}

The posterior correlations between the parameters could be studied by analyzing the shapes of 2D posterior distribution contours presented in the off-diagonal panels of Figs. \ref{fig1}-\ref{fig3}. Circular contours indicate that there is no correlation between a pair of the parameters or that it is very small, while stretched elliptical contours indicate a linear correlation (where a greater elongation corresponds to a greater correlation and a very narrow stretched contour designate a parameter degeneration). On the other hand, curved and irregularly shaped contours indicate more complex, non-linear correlations. Besides, the orientation of the contours from the lower left to the upper right corner corresponds to a positive correlation, while their slope from the upper left to the lower right corner indicate an anti-correlation.

Taking into account the above, and by comparing the results for the uniform priors against those for the Gaussian priors, it can be seen from Figs. \ref{fig1}-\ref{fig3} that, in all three studied cases, the uniform priors mostly induce complex, non-linear correlations, while the Gaussian priors give preference to the linear correlations. Among all the posterior  correlations, the most important for this study are those that do not correspond to the orbital elements, such as the correlations between the SMBH mass $M$ and a fifth force parameters $\delta$ and $\lambda$. One such correlation is between $\delta$ and $M$, which could appear due to the factor $\delta \cdot M / (1 + \delta)$ in the Yukawa-like correction given by the last term in the r.h.s of Eq. (\ref{eq:ppneom}) for EoM in extended PPN formalism. This factor corresponds to a rescaling of mass $M$ by $\delta$, and could potentially cause degeneration due to which it may be impossible to constrain these parameters separately. Another potentially significant correlation could be between a fifth force strength $\delta$ and its range $\lambda$, as indicated by their common increase (see Tables \ref{tab3}-\ref{tab5}). In order to investigate such potential correlations, we calculated the posterior correlation matrices for both types of the priors in all cases 1-3, extracted the relevant correlation coefficients and presented them in Table \ref{tab6}. As it can be seen from Table \ref{tab6}, most of the obtained results indicate little or no linear correlation, except in the case of the $M-\delta$ correlation. The $M-\delta$ correlation is noticeable for both types of the priors and in all three studied cases, but it is significantly higher in the case of the Gaussian priors, when it could reach $\sim 50\%$. Regarding a fifth force parameters $\delta$ and $\lambda$, no statistically significant linear correlation is found between them, which indicates that either these two parameters are not correlated after all, despite their common increase, or that there is some more complex non-linear correlation between them that could not be detected by the correlation coefficients. We believe it is the latter, since it was hinted in the respective panels of Figs. \ref{fig1}-\ref{fig3}, a similar non-linear correlation between $\delta$ and $\lambda$ was deduced in \cite{bork13}, and moreover it was shown that these two parameters should be related by the relationship (\ref{eq:lambda}), which will be further discussed in the following subsection.

\begin{table}[ht!]
\setlength{\tabcolsep}{4pt}
\centering
\caption[]{Correlation coefficients between posterior parameters $M$, $\delta$ and $\lambda$, extracted from respective full posterior correlation matrices.}
\label{tab6}
\begin{tabular}{r|r|r|r|r|r|r}
\hline
\noalign{\smallskip}
 & \multicolumn{3}{c|}{Uniform priors} &\multicolumn{3}{c}{Gaussian priors} \\
\noalign{\smallskip}
\cline{2-7}
\noalign{\smallskip}
& Case 1 & Case 2 & Case 3 & Case 1 & Case 2 & Case 3 \\
\noalign{\smallskip}
\hline
\noalign{\smallskip}
$M-\delta$       & -0.204 & -0.228 & -0.164 &  0.575 &  0.562 & 0.511 \\
$M-\lambda$      & -0.011 &  0.062 &  0.072 & -0.020 & -0.030 & 0.128 \\
$\lambda-\delta$ &  0.039 & -0.044 & -0.128 & -0.087 & -0.311 & 0.039 \\
\noalign{\smallskip}
\hline
\end{tabular}
\end{table}

\begin{table}[ht!]
\setlength{\tabcolsep}{4pt}
\centering
\caption[]{The range of a fifth force ($\lambda$) and its absolute errors ($\Delta\lambda$), estimated for 11 orders of magnitude of the strength ($\delta$), ranging from $\delta=10^{-5}$ to $\delta=10^{5}$. The necessary orbital elements and their uncertainties are taken from Table 3 of Ref. \cite{gill17}. The value of $f_{SP} = 1.10$, which was measured by the GRAVITY Collaboration \cite{abut20}, was used for calculating the absolute errors $\Delta\lambda$ according to expression (\ref{eq:lambdaerror}), assuming $\Delta\delta=0$ and the following three cases of the uncertainties $\Delta f_\mathrm{SP}$: $\pm 0.19$ (measured one), $\pm 0.1$ and $\pm 0.05$. The corresponding relative errors $\Delta\lambda/\lambda$ are 96.3\%, 51.3\% and 26.3\%, respectively.}
\label{tab7}
\begin{tabular}{p{1cm}|p{1.5cm}|p{2.5cm}p{2.5cm}p{2.5cm}}
\hline
\noalign{\smallskip}
$\delta$ & $\lambda$ (AU) &\multicolumn{3}{c}{$\Delta\lambda$ (AU)} \\
\noalign{\smallskip}
\cline{3-5}
\noalign{\smallskip}
& & $\Delta f_\mathrm{SP} = \pm 0.19$ & $\Delta f_\mathrm{SP} = \pm 0.10$ & $\Delta f_\mathrm{SP} = \pm 0.05$ \\
\noalign{\smallskip}
\hline
\noalign{\smallskip}
$10^{-5}$ &   209.9 & $\pm$   202.0 & $\pm$   107.6 &  $\pm$    55.1 \\
$10^{-4}$ &   663.6 & $\pm$   638.9 & $\pm$   340.3 &  $\pm$   174.4 \\
$10^{-3}$ &  2097.5 & $\pm$  2019.4 & $\pm$  1075.5 &  $\pm$   551.2 \\
$10^{-2}$ &  6603.2 & $\pm$  6357.4 & $\pm$  3385.9 &  $\pm$  1735.1 \\
$10^{-1}$ & 20008.7 & $\pm$ 19263.8 & $\pm$ 10259.8 &  $\pm$  5257.6 \\
        1 & 46924.6 & $\pm$ 45177.5 & $\pm$ 24061.4 &  $\pm$ 12330.3 \\
 $10^{1}$ & 63273.2 & $\pm$ 60917.4 & $\pm$ 32444.4 &  $\pm$ 16626.1 \\
 $10^{2}$ & 66032.1 & $\pm$ 63573.6 & $\pm$ 33859.1 &  $\pm$ 17351.1 \\
 $10^{3}$ & 66328.3 & $\pm$ 63858.7 & $\pm$ 34011.0 &  $\pm$ 17428.9 \\
 $10^{4}$ & 66358.2 & $\pm$ 63887.5 & $\pm$ 34026.3 &  $\pm$ 17436.8 \\
 $10^{5}$ & 66361.1 & $\pm$ 63890.3 & $\pm$ 34027.8 &  $\pm$ 17437.6 \\
\noalign{\smallskip}
\hline
\end{tabular}
\end{table}

\subsection{Potential small deviations from the GR prediction}

In order to check whether the results in all three studied cases are consistent with possible small deviations of S2 orbit from the GR prediction, we adopted the value of $f_{SP} = 1.10$, which was measured by the GRAVITY Collaboration \cite{abut20}, and used it to estimate the range of a fifth force ($\lambda$) and its absolute error ($\Delta\lambda$) for 11 orders of magnitude of the strength ($\delta$), ranging from $\delta=10^{-5}$ to $\delta=10^{5}$. For this purpose, we used the expressions (\ref{eq:lambda}) and (\ref{eq:lambdaerror}), assuming that $\Delta\delta=0$ and taking the necessary orbital elements and their uncertainties from Table 3 of Ref. \cite{gill17}. The obtained results are presented in Table \ref{tab7} for the following three uncertainties of $\Delta f_\mathrm{SP}$: $\pm 0.19$ (measured one), $\pm 0.1$ and $\pm 0.05$ (the latter two being our predictions for the future, more precise observations). These uncertainties correspond to the relative errors of $\Delta\lambda/\lambda\approx$ 96.3\%, 51.3\% and 26.3\%, respectively.

As it can be seen from Table \ref{tab7}, it is expected that the range $\lambda$ of a fifth force should be strongly affected by its strength $\delta$. $\lambda$ should start to increase with increase of $\delta$ until it reaches a certain threshold of $\lambda\sim$ 66000 AU for $\delta\sim 10$, after which it should remain nearly constant. Thus, these two parameters should be more correlated in the case of weaker interactions (e.g. for $\delta < 0.1$), but for much stronger interactions (e.g. for $\delta > 10$), $\delta$ and $\lambda$ should become almost independent (see Table \ref{tab7}).

By comparing the results for $\delta$ and $\lambda$, obtained by the MCMC analysis (see Tables \ref{tab3} and \ref{tab4}), with the corresponding estimates presented in Table \ref{tab7}, it can be seen that the results for all three cases are compatible, within the error intervals, with the measured value of $f_\mathrm{SP} = 1.10\pm 0.19$. Hence, future, more precise observations which will reduce uncertainty of $f_\mathrm{SP}$ to $\Delta f_\mathrm{SP}=\pm 0.1$, or even to $\Delta f_\mathrm{SP}=\pm 0.05$ are needed in order to obtain tighter constraints for a fifth force (see the corresponding columns in Table \ref{tab7}).

\section{Conclusions}

We investigated a possible presence of a fifth force at the GC and constrained its strength $\delta$ and range $\lambda$ by
studying its influence on the orbit of S2 star around the central SMBH of our Galaxy. For this purpose we simulated orbits of S2 star in the extended PPN formalism (\ref{eq:ppneom}) that predicts the emergence of a fifth force, and fitted the simulated orbits into the corresponding astrometric observations using MCMC method. The following three cases for a fifth force range $\lambda$ were studied: i) when $\lambda$ is about a few hundred AU (i.e. deep inside the orbit of S2 star), ii) when $\lambda$ is about a thousand AU (i.e. approximately the size of S2 star orbit), and iii) when $\lambda$ is several thousand AU (i.e. much larger than the size of S2 star orbit). We also studied whether the results in all three cases are consistent with possible small deviations of S2 orbit from the GR prediction. The obtained results can be summarized as follows:
\begin{enumerate}
\item
The strength $\delta$ increases and its relative error $\Delta\delta/\delta$ decreases with increase in the range $\lambda$;
\item
Although a significant linear correlation between the strength $\delta$ and the range $\lambda$ of a fifth force was not found, our additional theoretical considerations suggest that these two parameters could be correlated in a more complex, non-linear way, as well as that this non-linear correlation could be stronger in the case of weaker interactions (e.g. for $\delta < 0.1$), while these two parameters could be almost independent for much stronger interactions (e.g. for $\delta > 10$);
\item
The value $\delta\approx 0.005$, obtained in the first case, is in good agreement with the recent constraints on the strength and range of a fifth force obtained by the GRAVITY Collaboration \cite{abde25} for a similar length scale, but for a different Yukawa-like potential;
\item
The results obtained in the third case are compatible with our earliest estimates for Yukawa gravity parameters given in \cite{bork13};
\item
The strength $\delta$ and range $\lambda$ of a fifth force are almost model independent and different Yukawa-like potentials (as e.g. those given by the expressions (\ref{eq:pot1})-(\ref{eq:yukawapot})) could be used for modeling a fifth force;
\item
The obtained estimates in all three cases are compatible, within the error intervals, with the value of $f_\mathrm{SP} = 1.10\pm 0.19$ measured in the case of S2 star by GRAVITY Collaboration in 2020 \cite{abut20};
\item
The obtained uncertainties of a fifth force parameters are large, and thus we cannot exclude GR case. Therefore, future, more precise observations are needed in order to obtain tighter constraints on a fifth force.
\end{enumerate}

\authorcontributions{All coauthors participated in writing, calculation and discussion of obtained results.}

\funding{The Authors acknowledge the support of Ministry of Science, Technological Development and Innovations of the Republic of Serbia through the Project contracts No. 451-03-33/2026-03/200002 and 451-03-33/2026-03/200017.}

\dataavailability{Data are contained within the article.}

\conflictsofinterest{The authors declare no conflict of interest.}

\abbreviations{Abbreviations}{The following abbreviations are used in this manuscript:\\
\noindent
\begin{tabular}{@{}ll}
DM & Dark Matter \\
DE & Dark Energy \\
ETG & Extended Theories of Gravity \\
GC & Galactic Center \\
GR & General Relativity \\
GW & gravitational wave \\
MCMC simulation & Markov chain Monte Carlo simulation \\
PPN formalism & parameterized post-Newtonian formalism \\
SMBH & Supermassive black hole \\
\end{tabular}}

\reftitle{References}

\end{document}